# The Equations of Maxwell for a Medium in Prefered and Non-Prefered Reference Frames

Walter Petry
Mathematisches Institut der Universitaet Duesseldorf, D-40225 Duesseldorf
E-mail: wpetry@meduse.de
petryw@uni-duesseldorf.de

Abstract: Let us consider a reference frame for which the pseudo-Euclidean geometry is valid (prefered frame). The equations of Maxwell in empty space have a simple form and are derived from a Lagrangian. In a medium magnetic permeability and electric permittivity exist. The equations of Maxwell are also well-known in a medium but they cannot be derived as in empty space. In addition to the pseudo-Euclidean geometry a tensor of rank two is stated with which the proper time in a medium is defined. The theory of Maxwell now follows along the lines of the empty space. Reference frames which are uniformly moving with regard to the prefered frame which have an anisotropic geometry are studied by many authors and are called non-prefered reference frames. The equations of Maxwell of a medium in a non-prefered reference frame are derived by the use of the transformation formulae from the prefered to the non-prefered frame. A Lagrangian for the equations of Maxwell of a medium similar to the one in empty space is also given.

## 1. Introduction

Let us consider a reference frame $\Sigma'$ for which the pseudo-Euclidean geometry holds. The equations of Maxwell in empty space can be derived from a Lagrangean. This is a well-known result which can be found in many textbooks on Electrodynamics. In a medium there exist in addition to the electric and magnetic fields $E$ and $B$ the derived fields $D$ and $H$ through polarisation and magnetisation of the medium. It holds for a medium which is linear and isotropic

$$D = \varepsilon E, \quad B = \mu H. \tag{1.1}$$

Here, $\varepsilon$ is the constant permittivity and $\mu$ the constant magnetic permeability which characterize the medium. In the special case $\varepsilon = \mu = 1$ the medium corresponds to the empty space. The equations of Maxwell have the form

$$rotH - \frac{1}{c}\frac{\partial D}{\partial t} = \frac{4\pi}{c}J, \quad divD = 4\pi\rho \tag{1.2a}$$

$$rotE + \frac{1}{c}\frac{\partial B}{\partial t} = 0, \quad divB = 0 \tag{1.2b}$$

where $\rho$ and $J$ denote the electrical charge and the electrical current density. The equations of Maxwell (1.2) together with the relations (1.1) hold in the prefered reference frame $\Sigma'$. It is well-known that in a medium the velocity of light $c_L$ is different from the vacuum light velocity and it holds for the absolute value of the light velocity

$$c_L = c/\sqrt{\varepsilon\mu} = c/n \tag{1.3}$$

where $n$ is the index of refraction of the medium. In empty space the pseudo-Euclidean metric defines the proper time but in a medium the pseudo-Euclidean metric does not give the proper time because the light velocity in it differes from the vacuum light velocity. Therefore, in this e-print a tensor $(g_{ij}')$ is introduced containing the constants $\varepsilon$ and $\mu$. It implies the correct definition of the proper time. All these results are stated in the prefered reference frame $\Sigma'$ where the pseudo-Euclidean geometry is valid. By the use of the tensor $(g_{ij}')$ and its inverse a Lagrangean for Maxwell's equations is given similar to that in empty space. The equations of Maxwell in covariant form are derived. They are identical with the equations (1.2) together with (1.1).

It is worth to mention that Petry [1,2] has derived a theory of gravitation in flat space-time where in addition to the pseudo-Euclidean geometry (more generally a flat space-time geometry) a tensor $(g_{ij})$ is introduced to define the proper time for a moving object in the field $(g_{ij})$ which describes the gravitatioal field. Field equations for gravitation are derived. This theory is in good agreement with all the results of Einstein's general theory of relativity (see e.g. [3-6]).

A uniformly moving frame $\Sigma$ relative to the prefered frame is studied. The geometry of such a frame is anisotropic, i.e. the light velocity is direction dependent. There exist many studies about so-called non-prefered frames. This is quite different to Einstein's special theory of relativity where in any uniformly moving frame the pseudo-Euclidean geometry is valid with the well-known Lorentz-transformations. A new class of transformations from the prefered frame $\Sigma'$ to a uniformly moving non-prefered frame $\Sigma$ are given by Tangherlini [7]. Later on, these so-called Tangherlini-transformations have been studied by many authors (see e.g. [8-12]). Further remarks are found in paper [12] where also references about experiments of the two different conceptions are stated. Petry [10] started from the transformations

$$x^i = x^{i\,\prime} \ (i=1,2,3), \quad x^4 = x^{4\,\prime} - \left(x', \frac{v'}{c}\right) \tag{1.4a}$$

and the inverse formulae

$$x^{i\,\prime} = x^i \ (i=1,2,3), \quad x^{4\,\prime} = x^4 + \left(x, \frac{v'}{c}\right) \tag{1.4b}$$

where $(\cdot,\cdot)$ denotes the scalar product in $R^3$. Here, $x = (x^1, x^2, x^3)$ is the vector of the Cartesian coordinates and $x^4 = ct$. All the quantities with prime are measured in the prefered frame $\Sigma'$ and those without prime are the corresponding ones in the non-prefered frame $\Sigma$. The vector $v' = (v^{1\,\prime}, v^{2\,\prime}, v^{3\,\prime})$ is the velocity measured in $\Sigma'$ of the frame $\Sigma'$ relative to the frame $\Sigma$. An event at rest in $\Sigma'$ (resp. $\Sigma$) is transformed by the formulae (1.4) to the same event in $\Sigma$ (resp. $\Sigma'$) at rest. In contrast to the formulae (1.4), the Tangherlini-transformations

$$x^i = x^{i\,\prime} + (\gamma - 1)\frac{(x', v')}{|v'|^2} v^{i\,\prime} + \gamma \frac{v^{i\,\prime}}{c} x^{4\,\prime} \ (i=1,2,3), \quad x^4 = \gamma^{-1} x^{4\,\prime} \tag{1.5a}$$

where $|\cdot|$ denotes the Euclidean norm with the inverse formulae

$$x^{i\,\prime} = x^i + (\gamma^{-1} - 1)\frac{(x, v')}{|v'|^2} v^{i\,\prime} - \gamma \frac{v^{i\,\prime}}{c} x^4 \ (i=1,2,3), \quad x^{4\,\prime} = \gamma x^4 \tag{1.5b}$$

where

$$\gamma = \left(1 - \left|\frac{v'}{c}\right|^2\right)^{-1/2} \tag{1.5c}$$

give the result of an event in $\Sigma'$ (resp. $\Sigma$) as it is seen in $\Sigma$ (resp. $\Sigma'$). Hence, the meanings of the two formulae (1.4) and (1.5) are quite different from one another but in both cases the geometry in $\Sigma'$ (pseudo-Euclidean geometry) is transformed to the same anisotropic geometry in the non-prefered reference frame $\Sigma$ and conversely. Furthermore, the transformation formulae of the non-prefered frame $\Sigma$ which do not change the metric in $\Sigma$ are given. They correspond to the Lorentz-transformations in $\Sigma'$. In the non-prefered frame $\Sigma$ the equations of Maxwell in empty space and several other studies are given, as e.g. the results of Hoek, Fizeau and the Doppler effect. All these results are found in paper [10].

Subsequently, in addition to Maxwell's equations of a medium in the prefered frame $\Sigma'$ Maxwell's equations in a non-prefered frame $\Sigma$ are given. There are two possibilities: (1) the transformation of Maxwell's equations (1.1) with (1.2) in $\Sigma'$ to the corresponding ones in $\Sigma$ by the use of the transformations (1.4); (2) the introduction of a tensor $(g_{ij})$ in $\Sigma$ by the transformation of $(g_{ij}')$ in $\Sigma'$ by the use of (1.4) implying the equations of Maxwell of a medium in covariant form. The tensor $(g_{ij})$ contains the parameters $\varepsilon$ and $\mu$ of the medium and yields the definition of the proper time in $\Sigma$ of a moving body in the medium. With the aid of the tensor $(g_{ij})$ the equations of Maxwell in a medium may be written analogously to Maxwell's equations in empty space. The energy-momentum tensor of the electro-magnetic field and the conservation law of the total energy-momentum implying the equations of motion in a medium are given. As application in the non-prefered frame a plane electro-magnetic wave is studied.

It is worth to mention two books on Electrodynamics of Van Bladel [13] and of Hehl and Obukhov [14]. In both books accelerated reference frames are studied. In particular, the chapter on the metric by an alternative method in [14] is related to the present article.

## 2. Prefered Reference Frame

Let us consider a prefered reference frame $\Sigma'$ which is given by

$$(ds)^2 = -\eta_{ij}' dx^{i\,\prime} dx^{j\,\prime} \tag{2.1a}$$

with

$$(\eta_{ij}') = diag(1,1,1,-1). \tag{2.1b}$$

The inverse tensor has the form
$$(\eta^{ij\,\prime}) = diag(1,1,1,-1). \tag{2.2}$$

Subsequently, all the quantities in the prefered frame $\Sigma'$ have attached a prime to distinguish them from the corresponding quantities in the non-prefered frame $\Sigma$ without prime which will be studied in the next section. Let us follow along the lines of paper [10].

The light velocity in $\Sigma'$ is isotropic and always equal to the vacuum light velocity $c$. Let $w' = (w^{1\prime}, w^{2\prime}, w^{2\prime})$ be a constant velocity vector and put $\gamma = \left(1 - \left|\frac{w'}{c}\right|^2\right)^{-1/2}$. Then, the Lorentz-transformations have the form

$$\tilde{x}^{i\,\prime} = x^{i\,\prime} + (\gamma-1)\frac{(x',w')}{|w'|^2}w^{i\,\prime} + \gamma\frac{w^{i\,\prime}}{c}x^{4\,\prime} \quad (i=1,2,3), \quad \tilde{x}^{4\,\prime} = \gamma\left(x^{4\,\prime} + \left(x', \frac{w'}{c}\right)\right). \tag{2.3a}$$

The inverse formulae are given by

$$x^{i\,\prime} = \tilde{x}^{i\,\prime} + (\gamma-1)\frac{(\tilde{x}',w')}{|w'|^2}w^{i\,\prime} - \gamma\frac{w^{i\,\prime}}{c}\tilde{x}^{4\,\prime} \quad (i=1,2,3), \quad x^{4\,\prime} = \gamma\left(\tilde{x}^{4\,\prime} - \left(\tilde{x}', \frac{w'}{c}\right)\right). \tag{2.3b}$$

It is well-known that the Lorentz-transformations (2.3) do not change the line-element (2.1). Hence, knowing an event in the frame $\Sigma'$ described by the coordinates $(x^{i\,\prime})$ one obtains by the transformations (2.3) the same event also in the frame $\Sigma'$ with the coordinates $(\tilde{x}^{i\,\prime})$ if it is moving with constant velocity $w'$ relative to the first one. Therefore, Lorentz-transformations have nothing to do with a uniformly moving frame, i.e. the principle of special relativity is not assumed. The equations of Maxwell are well known and are studied in many textbooks. They are written in the form

$$rotH' - \frac{1}{c}\frac{\partial D'}{\partial t'} = \frac{4\pi}{c}J', \quad divD' = 4\pi\rho' \tag{2.4a}$$

$$rotE' + \frac{1}{c}\frac{\partial B'}{\partial t'} = 0, \quad divB' = 0 \tag{2.4b}$$

with the electrical current density $J' = (J^{1\prime}, J^{2\prime}, J^{3\prime})$ and the electrical charge density $\rho'$. In a simple medium with constant electrical permittivity $\varepsilon$ and constant magnetic permeability $\mu$ the connection are

$$D' = \varepsilon E', \quad B' = \mu H'. \tag{2.4c}$$

The special case $\varepsilon = \mu = 1$ implies no medium, i.e. empty space where simple covariant formulations of the equations of Maxwell are given by the use of the metric (2.1) (see e.g. [10]).
In the general case we define the tensor

$$(g_{ij}\,') = \sqrt{\mu}\,diag\left(1,1,1,-\frac{1}{\varepsilon\mu}\right) \tag{2.5a}$$

with the inverse tensor

$$(g^{ij\,\prime}) = \frac{1}{\sqrt{\mu}}\,diag(1,1,1,-\varepsilon\mu). \tag{2.5b}$$

The proper time in the medium is given by

$$c^2(d\tau)^2 = -g_{ij}\,' dx^{i\,\prime} dx^{j\,\prime} = \sqrt{\mu}\left(\frac{1}{\varepsilon\mu}(dct')^2 - |dx'|^2\right). \tag{2.6}$$

Hence, by $d\tau = 0$ we get the light velocity

$$\left|\frac{dx'}{dt'}\right| = \frac{c}{\sqrt{\varepsilon\mu}} = \frac{c}{n} \tag{2.7}$$

where $n = \sqrt{\varepsilon\mu}$ is the refraction index of the medium.

Such considerations in flat space-time and the introduction of a tensor $(g_{ij})$ and the proper time are studied in the papers [1,2]. Here, $(g_{ij})$ are the gravitational potentials which satisfy some nonlinear partial differential equations. This theory of gravitation in flat space-time is studied in several papers [1-6]. It gives all the results of general relativity but it implies non-singular cosmological models.

It is worth to mention that in a medium the introduction of the proper time by (2.1) with

$$(ds)^2 = c^2(d\tau)^2$$

is incorrect because $d\tau = 0$ implies the vacuum light velocity.

To get another form of Maxwell's equations in a medium we start from an antisymmetric tensor $F_{ij}'$ defined by

$$F_{ij}' = \frac{\partial A_j'}{\partial x^{i\prime}} - \frac{\partial A_i'}{\partial x^{j\prime}} \tag{2.8}$$

where $A_i'$ are the electro-magnetic potentials. Furthermore, let us define the antisymmetric tensor (in analogy to Electrodynamics in empty space)

$$F^{ij\prime} = g^{ik\prime} g^{jl\prime} F_{kl}' \tag{2.9}$$

and let $(J^{1\prime}, J^{2\prime}, J^{3\prime}, J^{4\prime})$ be the electrical four-current density. We consider the covariant differential equations in $\Sigma'$:

$$\frac{\partial}{\partial x^{k\prime}} F^{ki\prime} = \frac{4\pi}{c} J^{i\prime} \quad (i = 1,2,3,4) \tag{2.10a}$$

$$\frac{\partial}{\partial x^{k\prime}} F^{ij\prime} + \frac{\partial}{\partial x^{i\prime}} F^{jk\prime} + \frac{\partial}{\partial x^{j\prime}} F^{ki\prime} = 0 \quad (i, j, k = 1,2,3,4). \tag{2.10b}$$

It easily follows from the definition (2.8) that the differential equations (2.10b) are satisfied. The electrical field $E'$ and the magnetic field $B'$ are defined by

$$E' = (F_{41}', F_{42}', F_{43}'), \qquad B' = (F_{32}', F_{13}', F_{21}'), \tag{2.11}$$

then the differential equations (2.10b) are identical to the equations (2.4b) of Maxwell. Put for the derived fields

$$H' = (F^{32\prime}, F^{13\prime}, F^{21\prime}), \qquad D' = (F^{14\prime}, F^{24\prime}, F^{34\prime}), \tag{2.12}$$

then the differential equations (2.10a) are identical to the equations (2.4a) of Maxwell. Furthermore, it follows by (2.5b) and (2.9)

$$H' = \frac{1}{\mu} B', \qquad D' = \varepsilon E',$$

i.e. the relations (2.4c) are fulfilled. Hence, we have received a reformulation of the equations of Maxwell in a medium similar to Maxwell's equations in empty space (see e.g. [10]). Let us introduce the potentials $A_i'$ by relation (2.8) then the equations (2.10b) are fulfilled and $A_i'$ $(i = 1,2,3,4)$ must be calculated by the equations (2.10a). They are rewritten in the form

$$\frac{\partial}{\partial x^{m\prime}} g^{mk\prime} g^{il\prime} \left( \frac{\partial A_l'}{\partial x^{k\prime}} - \frac{\partial A_k'}{\partial x^{l\prime}} \right) = \frac{4\pi}{c} J^{i\prime} \quad (i = 1,2,3,4)$$

or equivalently

$$\frac{\partial}{\partial x^{m\prime}} g^{mk\prime} \left( \frac{\partial A_i'}{\partial x^{k\prime}} - \frac{\partial A_k'}{\partial x^{i\prime}} \right) = \frac{4\pi}{c} g_{ik}' J^{k\prime} \quad (i = 1,2,3,4).$$

With the aid of the Lorentz gauge

$$\frac{\partial}{\partial x^{m\prime}} (g^{mk\prime} A_k') = 0 \tag{2.13a}$$

the differential equations are equivalent to the equations

$$\frac{\partial}{\partial x^{m\prime}} \left( g^{mk\prime} \frac{\partial A_i'}{\partial x^{k\prime}} \right) = \frac{4\pi}{c} g_{ik}' J^{k\prime} \quad (i = 1,2,3,4). \tag{2.13b}$$

Hence, we have four differential equations (2.13b) with the gauge condition (2.13a) for the four potentials $A_i'$ $(i = 1,2,3,4)$.

The density of the Lagrangean of the electro-magnetic field is

$$L_E = -\frac{1}{4} F^{ij\prime} F_{ij}' + \frac{4\pi}{c} A_i' J^{i\prime} \tag{2.14}$$

with (2.8) and (2.9). The energy-momentum tensor of the electro-magnetic field has the form

$$T^i{}_j(E)' = \frac{1}{4\pi}\left(F^{ik}{}'F_{jk}{}' - \frac{1}{4}\delta^i{}_j F^{kl}{}'F_{kl}{}'\right). \tag{2.15a}$$

It easily follows that the tensor

$$T^{ij}(E)' = g^{jk}{}' T^i{}_k(E)' \tag{2.15b}$$

is symmetric.
The equations of motion of a charged particle follows from the consevation law of the total energy- momentum:

$$\frac{\partial}{\partial x^{k'}}\left(T^k{}_i(E)' + T^k{}_i(M)'\right) = 0 \quad (i=1,2,3,4) \tag{2.16}$$

with the energy-momentum tensor of matter

$$T^i{}_j(M)' = \rho_m{}' g_{jk}{}' \frac{dx^{k'}}{d\tau}\frac{dx^{i'}}{d\tau} \tag{2.17}$$

where $\rho_m{}'$ denotes the density of matter.

### 3. Non-Prefered Reference Frame

In this section Maxwell's equations of a medium in a non-prefered reference frame $\Sigma$ are derived. Let us start from Maxwell's equations of a medium in the prefered frame $\Sigma'$, i.e. (2.4), (2.11) and (2.12). To receive the corresponding equations in the non-prefered frame we must consider the transformations (1.4) and not the Tangherlini-transformations (1.5). In the frame $\Sigma$ the fields are analogously defined as in the prefered frame $\Sigma'$ (see (2.11) and (2.12)), i.e.

$$E = (F_{41}, F_{42}, F_{43}), \quad B = (F_{32}, F_{13}, F_{21}), \quad H = (F^{32}, F^{13}, F^{21}), \quad D = (F^{14}, F^{24}, F^{34}). \tag{3.1}$$

Then, we get by the use of (1.4) and standard transformation rules

$$E = E', \quad H = H', \quad B = B' - \frac{v'}{c}\times E = \mu H - \frac{v'}{c}\times E, \quad D = D' + \frac{v'}{c}\times H = \varepsilon E + \frac{v'}{c}\times H \tag{3.2}$$

where in the last two relations the connections (2.4c) are used. By longer elementary calculations the equations of Maxwell (2.4a) and (2.4b) together with the relations (3.2) are transformed to the equations in the non-prefered frame $\Sigma$:

$$rotH - \frac{1}{c}\frac{\partial}{\partial t}\left(\varepsilon E + \frac{v'}{c}\times H\right) = J, \quad \varepsilon\left(divE - \frac{1}{c}\frac{\partial}{\partial t}\left(\frac{v'}{c},E\right)\right) = J^4 + \left(\frac{v'}{c},J\right)$$

$$rotE + \frac{1}{c}\frac{\partial}{\partial t}\left(\mu H - \frac{v'}{c}\times E\right) = 0, \quad \mu\left(divH - \frac{1}{c}\frac{\partial}{\partial t}\left(\frac{v'}{c},H\right)\right) = 0. \tag{3.3}$$

It follows by scalar multiplication of the first equation and the third one with $\frac{v'}{c}$

$$\left(\frac{v'}{c},J\right) + \frac{1}{c}\varepsilon\frac{\partial}{\partial t}\left(\frac{v'}{c},E\right) = \left(\frac{v'}{c},rotH\right) = -div\left(\frac{v'}{c}\times H\right)$$

$$\frac{1}{c}\mu\frac{\partial}{\partial t}\left(\frac{v'}{c},H\right) = -\left(\frac{v'}{c},rotE\right) = div\left(\frac{v'}{c}\times E\right).$$

These relations together with (3.2) imply by the use of (3.3) the equations of Maxwell in the non-prefered reference frame $\Sigma$:

$$rotH - \frac{1}{c}\frac{\partial}{\partial t}D = J, \quad divD = J^4, \tag{3.4a}$$

$$rotE + \frac{1}{c}\frac{\partial}{\partial t}B = 0, \quad divB = 0 \tag{3.4b}$$

with the relations (see (3.2))

$$B = \mu H - \frac{v'}{c}\times E, \quad D = \varepsilon E + \frac{v'}{c}\times H. \tag{3.4c}$$

The equations (3.4a) and (3.4b) are the well-known equations of Maxwell whereas the connections of the fields are now given by (3.4c).
Let us now consider the new formulation of Maxwell's equations by the use of (2.5) and (2.6). Again, the transformations (1.4) must be used. It follows for the medium in $\Sigma$:

$$g_{ij} = \sqrt{\mu}\left(\delta_{ij} - \frac{1}{\varepsilon\mu}\frac{v^{i\prime}}{c}\frac{v^{j\prime}}{c}\right) \qquad i,j = 1,2,3$$

$$= -\sqrt{\mu}\frac{1}{\varepsilon\mu}\frac{v^{i\prime}}{c} \qquad i = 1,2,3; j = 4$$

$$= -\sqrt{\mu}\frac{1}{\varepsilon\mu}\frac{v^{j\prime}}{c} \qquad i = 4; j = 1,2,3$$

$$= -\sqrt{\mu}\frac{1}{\varepsilon\mu} \qquad i = j = 4 \tag{3.5a}$$

with the inverse

$$g^{ij} = \frac{1}{\sqrt{\mu}}\delta^{ij} \qquad i,j = 1,2,3$$

$$= -\frac{1}{\sqrt{\mu}}\frac{v^{i\prime}}{c} \qquad i = 1,2,3; j = 4$$

$$= -\frac{1}{\sqrt{\mu}}\frac{v^{j\prime}}{c} \qquad i = 4; j = 1,2,3$$

$$= -\frac{1}{\sqrt{\mu}}\left(\varepsilon\mu - \left|\frac{v'}{c}\right|^2\right) \qquad i = j = 4 \tag{3.5b}$$

with the proper time

$$c^2(d\tau)^2 = -g_{ij}dx^i dx^j. \tag{3.6}$$

The metric $(\eta_{ij})$ in $\Sigma$ is given in paper [10] but it also follows by the relations (3.5) with $\varepsilon = \mu = 1$.
Relation (3.6) together with (3.5a) gives the absolute value of the light velocity

$$\left|\frac{dx_L}{dt}\right| = \frac{c}{\sqrt{\varepsilon\mu}}\bigg/\left(1 - \frac{1}{\sqrt{\varepsilon\mu}}\left|\frac{v'}{c}\right|\cos(v';v_L)\right) \tag{3.7}$$

where $(v'; v_L)$ denotes the angle between the velocity $v'$ and the light velocity $v_L$. The equations of Maxwell can be rewritten in the covariant form:

$$\frac{\partial}{\partial x^k}F^{ki} = \frac{4\pi}{c}J^i \quad (i = 1,2,3,4) \tag{3.8a}$$

where $J^i$ ($i = 1,2,3,4$) is the electrical four-current and

$$\frac{\partial}{\partial x^k}F_{ij} + \frac{\partial}{\partial x^i}F_{jk} + \frac{\partial}{\partial x^j}F_{ki} = 0 \quad (i,j,k = 1,2,3,4). \tag{3.8b}$$

Here, it holds again

$$F^{ij} = g^{ik}g^{jl}F_{kl} \quad (i,j = 1,2,3,4) \tag{3.8c}$$

with the relation

$$F_{ij} = \frac{\partial A_j}{\partial x^i} - \frac{\partial A_i}{\partial x^j} \quad (i,j = 1,2,3,4). \tag{3.8d}$$

By vitue of (3.8d) the relation (3.8b) is fulfilled. With the identification (3.1) the equations (3.8a) and (3.8b) are Maxwell's equations (3.4a) and (3.4b). The relations (3.8c) give by the use of (3.5b) and the definitions (3.1) the relations (3.4c). Hence, the equations (3.8) are identical with Maxwll's equations (3.4) in a medium in the non-prefered reference frame $\Sigma$. Furthermore, all the results of (2.13) to (2.17) in $\Sigma'$ are also valid in the non-prefered frame $\Sigma$ and they have the same form omitting the prime.

### 4. Plane Electro-Magnetic Wave
In this section the equations of Maxwell are applied to electro-magnetic waves in a medium. Let us consider the eqations (3.3) with (3.4c). It holds for waves

$$J^i = 0 \quad (i = 1,2,3,4). \tag{4.1}$$

Let $k$ be the wave vector and $k_4 = \pm 2\pi \upsilon / c$ where $\upsilon$ denotes the frequency. Put

$$E = E_0 \exp(i((k,x) + k_4 x^4)), \quad H = H_0 \exp(i((k,x) + k_4 x^4))$$
$$D = D_0 \exp(i((k,x) + k_4 x^4)), \quad B = B_0 \exp(i((k,x) + k_4 x^4)) \tag{4.2}$$

where $E_0, H_0, D_0$ and $B_0$ are fixed vectors. The equations (3.3) with (3.4c) give

$$k \times H_0 - k_4 D_0 = 0, \quad (k, E_0) - k_4 \left(\frac{v'}{c}, E_0\right) = 0,$$

$$k \times E_0 + k_4 B_0 = 0, \quad (k, H_0) - k_4 \left(\frac{v'}{c}, H_0\right) = 0. \tag{4.3}$$

The second and the fourth equation of (4.3) imply

$$\left(k - k_4 \frac{v'}{c}, E_0\right) = 0, \quad \left(k - k_4 \frac{v'}{c}, H_0\right) = 0. \tag{4.4}$$

Vector multiplication of the first equation (4.3) with $H_0$ and the third one with $E_0$ yield by the use of (4.4)

$$k - k_4 \varepsilon \frac{H_0 \times E_0}{|H_0|^2} - k_4 \frac{v'}{c} = 0,$$

$$k - k_4 \mu \frac{H_0 \times E_0}{|E_0|^2} - k_4 \frac{v'}{c} = 0. \tag{4.5}$$

These two relations give

$$\varepsilon |E_0|^2 = \mu |H_0|^2. \tag{4.6}$$

It follows by inserting the first equation (4.5) in the first equation (4.3) and the use of (3.4c)

$$(H_0, E_0) = 0. \tag{4.7}$$

Relation (4.5) gives

$$|k|^2 = \left(k_4 \frac{v'}{c} + k_4 \mu \frac{H_0 \times E_0}{|E_0|^2}, k_4 \frac{v'}{c} + k_4 \mu \frac{H_0 \times E_0}{|E_0|^2}\right) =$$

$$= k_4^2 \left(\left|\frac{v'}{c}\right|^2 + 2 \frac{\mu}{|E_0|^2}\left(\frac{v'}{c}, H_0 \times E_0\right) + \mu^2 \frac{|H_0 \times E_0|^2}{|E_0|^4}\right). \tag{4.8}$$

Multiplication of (4.5) with $\dfrac{v'}{c}$ implies

$$k_4 \frac{\mu}{|E_0|^2}\left(\frac{v'}{c}, H_0 \times E_0\right) = \left(\frac{v'}{c}, k\right) - k_4 \left|\frac{v'}{c}\right|^2.$$

The last expression of relation (4.8) is written by the use of (4.7) and (4.6)

$$\mu^2 \frac{|H_0 \times E_0|^2}{|E_0|^4} = \mu^2 \frac{|H_0|^2 |E_0|^2 \sin^2(E_0; H_0)}{|E_0|^4} = \mu^2 \frac{|H_0|^2}{|E_0|^2} = \varepsilon \mu.$$

Then, (4.8) is rewritten by the use of the last two relations

$$|k|^2 = k_4^2 \left(\varepsilon \mu - \left|\frac{v'}{c}\right|^2\right) + 2k_4 \left(k, \frac{v'}{c}\right).$$

Hence, we have

$$k_4^2 \left(\varepsilon \mu - \left|\frac{v'}{c}\right|^2\right) + 2k_4 |k|\left|\frac{v'}{c}\right|\cos(k; v') - |k|^2 = 0.$$

The solution of this quadratic equation is given by

$$k_4 = \frac{|k|}{\sqrt{\varepsilon\mu}}\left(\pm\sqrt{1-\left|\frac{1}{\sqrt{\varepsilon\mu}}\frac{v'}{c}\right|^2 \sin^2(k;v')} - \left|\frac{1}{\sqrt{\varepsilon\mu}}\frac{v'}{c}\right|\cos(k;v')\right) \bigg/ \left(1-\left|\frac{1}{\sqrt{\varepsilon\mu}}\frac{v'}{c}\right|^2\right) \qquad (4.9)$$

where the upper or the lower sign holds. Hence, it follows by the use of $k_4 = \pm 2\pi\upsilon/c$ for the frequency $\upsilon$ and the Euclidean norm of the wave vector $|k|$:

$$\frac{2\pi\upsilon}{|k|} = \frac{c}{\sqrt{\varepsilon\mu}} \bigg/ \left(\sqrt{1-\left|\frac{1}{\sqrt{\varepsilon\mu}}\frac{v'}{c}\right|^2 \sin^2(k;v')} \pm \left|\frac{1}{\sqrt{\varepsilon\mu}}\frac{v'}{c}\right|\cos(k;v')\right). \qquad (4.10)$$